\documentclass[10pt, conference, compsocconf]{IEEEtran} 
\usepackage{listings}
\usepackage{color}
\definecolor{light-gray}{gray}{0.80}
\usepackage{url}
\usepackage{graphicx}
\usepackage{cite}

\begin{document}

\title{A Survey of the State of the Art in Data Mining and Integration Query Languages}

\author{\IEEEauthorblockN{Sabri Pllana, Ivan Janciak, Peter Brezany, and Alexander W\"{o}hrer}
\IEEEauthorblockA{Research Group of Scientific Computing\\
University of Vienna\\
Nordbergstrasse 15/C/3, 1090 Vienna, Austria\\
\{pllana,janciak,brezany,woehrer\}@par.univie.ac.at
}}

\IEEEspecialpapernotice{\footnotesize(2011 International Conference on Network-Based Information Systems)}

\maketitle

%\thispagestyle{empty}

%------------------------------------------------------------------------- 
\begin{abstract}
The major aim of this survey is to identify the strengths and weaknesses of a representative set of Data-Mining and Integration (DMI) query languages. We describe a set of properties of DMI-related languages that we use for a systematic evaluation of these languages. In addition, we introduce a scoring system that we use to quantify our opinion on how well a DMI-related language supports a property. The languages surveyed in this paper include: DMQL, MineSQL, MSQL, M2MQL, dmFSQL, OLEDB for DM, MINE RULE, and Oracle Data Mining. This survey may help researchers to propose a DMI language that is beyond the state-of-the-art, or it may help practitioners to select an existing language that fits well a purpose.

\end{abstract}


%------------------------------------------------------------------------- 
\section{Introduction}

Data integration, the process of combining data from different sources, has been an active research topic for many years~\cite{17}. A perfectly integrated information system would give the illusion that users interact with only one central, locally running, homogeneous and consistent information system providing location-, schema- and language transparency~\cite{17}. Enabled by the network performance improvement, data integration research promises to have similar effects as data mining~\cite{15}. 

Data Mining (DM), also popularly referred to as Knowledge Discovery in Databases (KDD), is the automated or convenient extraction of patterns representing knowledge implicitly stored in large volumes of data~\cite{18,khan2010,GridMiner2008}. In recent years there has been a rapid growth in data mining techniques~\cite{19}. Data mining has two main high-level goals: prediction and description. Commonly used methods for data mining are: association rules, sequential patterns, classification, regression, clustering, and change and deviation detection. From user\textquoteright s point of view, the execution of a data mining process and the discovery of a set of patterns can be considered either as an answer to a sophisticated database query or as a result of an execution of a data mining workflow. The first is called the \emph{descriptive approach}, while the second is the \emph{procedural approach}.

According to the CRoss Industry Standard Process for Data Mining (CRISP-DM)~\cite{28}, the life-cycle of a data mining project consists of six phases. These phases include: \emph{business understanding} (understanding the project objectives and requirements); \emph{data understanding} (get familiar with the data and identify data quality problems); \emph{data preparation} (also known as data pre-processing, is about how to construct the final dataset); \emph{modelling} (data mining techniques are selected and applied and their parameters are calibrated); \emph{evaluation} (review the steps executed to construct the model to be certain it properly achieves the business objectives); \emph{deployment} (what actions will need to be carried out in order to actually make use of the created models).

On-Line Analytical Processing (OLAP) is an approach for providing fast answers to analytical queries that are dimensional in nature. Systems configured for OLAP often employ a multidimensional data model, also called a cube, allowing for complex analytical and ad-hoc queries with a rapid execution time. 

The DMI query languages developed so far support the above mentioned functionality
at different level, as discussed in this survey.

%------------------------------------------------------------------------- 
\subsection{Related Work}

For obvious reasons commercially-available surveys are not discussed here. Academic surveys related to our work are scarce. A survey of data mining query languages for the domain of multimedia databases is presented in~\cite{mustafa05}. In~\cite{goebel99} authors present a survey of 43 data mining tools. In contrast to the related work our survey is not limited to a particular domain (such as multimedia) or to data mining tools, but it considers both \emph{data mining} and \emph{integration} aspects of a representative set of languages. 

%------------------------------------------------------------------------- 
\subsection{Contributions}

In this paper we systematically evaluate a representative set of state-of-the-art DMI-related languages: (1) DMQL, (2) MineSQL, (3) MSQL, (4) M2MQL, (5) dmFSQL, (6) OLE-DB for DM, (7) MINE RULE, and (8) Oracle Data Mining. For each language we provide an overview and we state the interesting features and drawbacks that we have observed. 

For the purpose of systematic language evaluation, we have developed the following list of properties (or criteria): data integration, data pre-processing, OLAP, presentation, DM methods, implementation, granularity, model processing, CRISP-DM compliance, background knowledge, and extensibility. For each property we provide a single paragraph description on how well the language supports the property. We use our scoring system introduced in Section~\ref{sec:approach} to quantify our opinion on how well a DMI-related language supports selected properties: insufficient, poor, sufficient, good, very good, excellent. 

The rest of this paper is structured as follows. The list of properties and the scoring system is described in Section~\ref{sec:approach}. In Section~\ref{sec:survey} we systematically evaluate the selected DMI-related languages. The results of the evaluation are summarised in Section~\ref{sec:summary}. Section~\ref{sec:conclusions} concludes this paper. 

%------------------------------------------------------------------------- 
\section{Evaluation Approach}
\label{sec:approach}

In this section we describe a set of properties of DMI-related languages. We use these properties for our subsequent systematic evaluation of DMI-related languages. In addition, we introduce the scoring system that we use to quantify our opinion on how well a DMI-related language supports a property. 

Our intention was to develop a comprehensive but manageable set of properties. The following properties are used in our survey:

\begin{itemize}
\item[p1] \textbf{Data integration}. Indicates whether the language provides constructs for integration of data from various data sources that may be geographically distributed. 
\item[p2] \textbf{Data pre-processing}. Indicates whether the language provides constructs for data pre-processing (such as selection, cleaning, or transformation).
\item[p3] \textbf{OLAP}. Indicates whether the language provides constructs for OLAP.
\item[p4] \textbf{Presentation}. Indicates whether the language supports visualization of results, and specification of the output format (or type). 
\item[p5] \textbf{DM methods}. Indicates the DM methods that are supported.
\item[p6] \textbf{Implementation}. Indicates whether and how well the language was implemented. In addition, it indicates the degree of the implementation portability. 
\item[p7] \textbf{Granularity}. Indicates the specification level (such as operator-level, algorithmic-level, or task level) and parameterization capabilities.
\item[p8] \textbf{Model processing}. Indicates the support for model verification, evaluation, validation, visualization, or reuse.
\item[p9] \textbf{CRISP-DM compliance}. Indicates the support for phases and tasks of the CRISP-DM reference model. 
\item[p10] \textbf{Background knowledge}. Indicates the support for concept hierarchies and type mapping needed by various data mining techniques. 
\item[p11] \textbf{Extensibility}. Indicates the easiness of extension with new language constructs.
\end{itemize}

The following grades (from 0 to 5) are used in our scoring system for quantifying how well a DMI-related language supports a property:

\begin{itemize}
\item	\textbf{Insufficient} (0). Indicates that the property is not supported at all.
\item	\textbf{Poor} (1). Indicates that there is a little support for the property; it only scratches the surface.
\item	\textbf{Sufficient} (2). Indicates that there is a sufficient support for the property, but many relevant improvements are possible. 
\item	\textbf{Good} (3). Indicates that there is a good support for the property, but some obvious improvements could be made. 
\item	\textbf{Very good} (4). Indicates that there is a very good support for the property, but some small improvements are possible.
\item	\textbf{Excellent} (5). Indicates that there is an excellent support for the property. To the best of our knowledge, nothing relevant could be added. 
\end{itemize}

%------------------------------------------------------------------------- 
\section{The Survey}
\label{sec:survey}

In this section we systematically evaluate a set of state-of-the-art DMI-related languages. The following languages are considered in our survey: (1) DMQL, (2) MineSQL, (3) MSQL, (4) M2MQL, (5) dmFSQL, (6) OLE-DB for DM, (7) MINE RULE, and (8) Oracle Data Mining.

%------------------------------------------------------------------------- 
\subsection{L1: DMQL}

The DMQL language was designed within the DBMiner~\cite{2} project for mining several kinds of knowledge in relational databases. The language allows the user to specify templates of rules to be discovered, called metapatterns. These metapatterns can be used to impose strong syntactic constraints on rules to be discovered. Portions of the language have been implemented in the DBMiner system for interactive mining of multiple-level knowledge in relational databases~\cite{1}.

The language consists of the specifications of four major primitives that should be specified in a data mining request:

\begin{enumerate}
\item Set of data in relevance to a data mining process
\item Kind of knowledge to be discovered 
\item Background knowledge (e.g., concept hierarchy)
\item Justification of the interestingness of the knowledge (i.e., thresholds)
\end{enumerate}

Additionally, the language provides syntax for five different data mining patterns: generalized relations, characteristic rules, discriminant rules, classification rules, and association rules. 

\begin{itemize}
\item [] \textbf{Evaluation of Properties}
\item[p1] \emph{Data integration}. The language does not provide any constructs specific for data integration. Although the language is based on the SQL syntax, the constructs such as JOIN are supported only for local data-sets.
\item[p2] \emph{Data pre-processing}. The language does not support any constructs specifically for data cleaning and transformation, but data selection is supported by the FROM, WHERE and HAVING constructs that are allowed as constraints for data selection queries.
\item[p3] \emph{OLAP}. OLAP is not supported.  The OLAP specific functions like roll up and drill down are used in the language but have different meaning because they can be only applied on the mined rules to traverse attribute hierarchies.
\item[p4] \emph{Presentation}. The language uses a display as a construct to specify presentation of the discovered patterns in forms including rules, tables, crosstabs, pie, bar chart, decision trees, cubes, curves, or surfaces.
\item[p5] \emph{DM methods}. The language provides specifications for the following data mining methods: generalized relations, characteristic rules, discriminant rules, classification rules, and association rules.
\item[p6] \emph{Implementation}. The language has been fully implemented in the  DBMiner system.
\item[p7] \emph{Granularity}. The language operates on the level of data mining methods and allows users only limited possibilities to specify some threshold parameters.
\item[p8] \emph{Model processing}. The language provides some constructs for model visualization.
\item[p9] \emph{CRISP DM compliance}. The language supports only the modelling phase of the CRISP-DM reference model.
\item[p10] \emph{Background knowledge}. The language uses concept hierarchies in order to express mined knowledge at different levels.
\item[p11] \emph{Extensibility}. DMQL can be easily extended by standard SQL constructs or by data mining methods.
\end{itemize}

\begin{table}[h]
\begin{center}
\begin{tabular}{|c|c|c|c|c|c|c|c|c|c|c|}
\hline
{\bf p1} & {\bf p2} & {\bf p3} & {\bf p4} & {\bf p5} & {\bf p6} & {\bf p7} & {\bf p8} & {\bf p9} & {\bf p10} & {\bf p11} \\
\hline
1 & 2 & 0 & 4 & 2 & 5 & 2 & 1 & 1 & 2 & 4 \\
\hline
\end{tabular} 
\end{center}
\caption{Summary of evaluation of properties for DMQL.} 
\label{tab:DMQL}
\end{table}

%------------------------------------------------------------------------- 
\subsection{L2: MineSQL}

MineSQL is a declarative, SQL like language developed at the Institute of Computing Science, Poznan University of Technology. MineSQL is an extension of SQL and serves as a uniform Application Programming Interface (API)  for building business applications dealing with knowledge discovery~\cite{5}. Association, characteristic, classification and discriminant rules are supported. Using these techniques, MineSQL enables the user to express rule queries. Moreover, MineSQL provides support for conceptual hierarchies and generalized and multiple level rules. Rule discovery, storage and manipulation are supported as well. For storing generated rules in relational databases it defines a new data type: the RULE. The rules can be integrated into existing databases and are searchable via a single query. The RULE data type supports various functions like support(), confidence(), body(), etc; which provide access to the rule elements as well as facilitate the conversion of values between RULE and other data types. To generate rules, MineSQL introduces a data discovery statement called MINE. Moreover, the MINE statement can be used as a sub-query in another statement, e.g. SELECT, or INSERT. 

Some of the interesting features on MineSQL are:
\begin{itemize}
\item Support for storing rules for further use in the same database.
\item The ability to search the rules as well as the data in one query.
\item The MINE statement can be used as a sub-query.
\end{itemize}

Limitations of MineSQL include:
\begin{itemize}
\item Support only for a limited set of data mining techniques.
\item Rule discovery/querying issues are not adequately addressed.
\end{itemize}

\begin{itemize}
\item [] \textbf{Evaluation of Properties}
\item[p1] \emph{Data integration}. There is no evidence in the studied literature that MineSQL supports data integration, although the integration of local data-sets via JOIN is supported.
\item[p2] \emph{Data pre-processing}. The data pre-processing steps like selection, cleaning and transformation are partially supported; MineSQL has strong support for data selection through the FROM and WHERE clauses of MINE statements but has little or no support for cleaning and transformation.
\item[p3] \emph{OLAP}. MineSQL does not provide constructs for OLAP.
\item[p4] \emph{Presentation}. MineSQL only displays the generated rules but has no support for visualization of results and/or specification of output format.
\item[p5] \emph{DM methods}. A basic set of DM methods i.e. association rule, characteristic rule, classification rule, conceptual hierarchy and generalization rule; are supported and enable the user to express his/her queries. 
\item[p6] \emph{Implementation}. A research prototype system has been implemented.
\item[p7] \emph{Granularity}.	A user can specify requirements at the task level. MineSQL provides FOR and TO clauses of MINE statements through which the user can specify the body and head of rule he/she is interested in.
\item[p8] \emph{Model processing}.	Model processing is partially supported.
\item[p9] \emph{CRISP DM compliance}.	Data preparation and data modeling are supported.
\item[p10] \emph{Background knowledge}.	MineSQL allows the definition and construction of taxonomy and generalized multiple level and multi-attribute association rules.
\item[p11] \emph{Extensibility}. MineSQL is developed from SQL, and therefore may be easily extensible with new constructs.
\end{itemize}

\begin{table}[h]
\begin{center}
\begin{tabular}{|c|c|c|c|c|c|c|c|c|c|c|}
\hline
{\bf p1} & {\bf p2} & {\bf p3} & {\bf p4} & {\bf p5} & {\bf p6} & {\bf p7} & {\bf p8} & {\bf p9} & {\bf p10} & {\bf p11} \\
\hline
1 & 3 & 0 & 1 & 3 & 2 & 3 & 1 & 1 & 3 & 3 \\
\hline
\end{tabular} 
\end{center}
\caption{Summary of evaluation of properties for MineSQL.} 
\label{tab:MineSQL}
\end{table}

%------------------------------------------------------------------------- 
\subsection{L3: MSQL}

MSQL was designed by Imielinski and Virmani in 1999~\cite{7}. This language supports association rule-mining. The language has the following properties~\cite{9}:

\begin{itemize}
\item It is an extension of SQL.
\item It supports the closure property in which saved results of the queries can be query.
\item It supports cross-over operations between data and rules i.e. the rules can be mapped back to the source data from which these are first extracted.
\item	The language can generate rules and query those rules using the same syntax.
\end{itemize}

MSQL rule generation is based on descriptors~\cite{8}. A descriptor is basically a pair of attribute names and values. MSQL defines support and confidence thresholds with which to retrieve the rule. The resultant rule body can consist of multiple attributes, while its head will have only one attribute. The syntax of this language supports a set of new data mining operators. For instance, the operator GETRULES generates the association rules. 

\begin{itemize}
\item [] \textbf{Evaluation of Properties}
\item[p1] \emph{Data integration}. There is no evidence in the studied literature that MSQL supports data integration.
\item[p2] \emph{Data pre-processing}.	It has partial support for the pre-processing, i.e. it can convert continuous data into discrete data, but it lacks other pre-processing tasks like sampling.
\item[p3] \emph{OLAP}. There is no support for OLAP.
\item[p4] \emph{Presentation}.	MSQL offers some post processing primitives using the operator SelectRules. 
\item[p5] \emph{DM methods}.	MSQL only supports association rule mining.
\item[p6] \emph{Implementation}.	MSQL was designed by Imielinski and Virmani in 1999 at the Rutgers University, New Jersey, USA. It is an academic project; we found no information on implementation.
\item[p7] \emph{Granularity}.	The specification level of MSQL is operator level. It supports operators such as GetRules or SelectRules.
\item[p8] \emph{Model processing}.	There is no support for model processing.
\item[p9] \emph{CRISP DM compliance}.	It supports data modelling.
\item[p10] \emph{Background knowledge}.	There is no support for background knowledge.
\item[p11] \emph{Extensibility}.	The language is designed only for association rules, so it is difficult (if not impossible) to accommodate other models for extending the language.
\end{itemize}

\begin{table}[h]
\begin{center}
\begin{tabular}{|c|c|c|c|c|c|c|c|c|c|c|}
\hline
{\bf p1} & {\bf p2} & {\bf p3} & {\bf p4} & {\bf p5} & {\bf p6} & {\bf p7} & {\bf p8} & {\bf p9} & {\bf p10} & {\bf p11} \\
\hline
0 & 2 & 0 & 1 & 1 & 2 & 3 & 0 & 1 & 0 & 1 \\
\hline
\end{tabular} 
\end{center}
\caption{Summary of evaluation of properties for MSQL.} 
\label{tab:MSQL}
\end{table}

%------------------------------------------------------------------------- 
\subsection{L4: M2MQL}

M2MQL~\cite{10} is a declarative language and it is compatible with SQL (it is an extension of SQL). It focuses on nested relational algebra for integrating multi-step association rule mining. It is based on the formal semantics sustained by the nested relational algebraic operations. M2MQL supports association rule mining primitives, which provide users with the flexibility to reuse the mining results and to conduct mining tasks based on them.

M2MQL supports the following kinds of association rules:

\begin{itemize}
\item Simple item-based association rules, such as rules that satisfy the minimal thresholds.
\item Generalized item-based association rules.
\item Quantitative association rules.
\end{itemize}

The language syntax provides CREATE RULE PRIMITIVE statement that is used for rule type-declaration, rule schema-illustration and rule generation. The SELECT RULE PRIMITIVE statement provides interactive processing of generated rule sets and original data sets, thereby supporting the reuse of existing rule sets in other data mining queries.

\begin{itemize}
\item [] \textbf{Evaluation of Properties}
\item[p1] \emph{Data integration}. There is no evidence in the studied literature that M2MQL supports data integration.
\item[p2] \emph{Data pre-processing}.	There is no support for pre-processing.
\item[p3] \emph{OLAP}. There is no support for OLAP.
\item[p4] \emph{Presentation}. It provides select and project operators to view the result of mined rules.
\item[p5] \emph{DM methods}. It only supports association rule mining.
\item[p6] \emph{Implementation}.	There is no implementation.
\item[p7] \emph{Granularity}.	The specification level of M2MQL is at operator level.
\item[p8] \emph{Model processing}. There is no support.
\item[p9]	\emph{CRISP DM compliance}. It supports the data modelling phase.
\item[p10] \emph{Background knowledge}.	There is no support for background knowledge.
\item[p11] \emph{Extensibility}.	The language is designed only for association rules, so it is difficult to extend the language with other kinds of models.
\end{itemize}

\begin{table}[h]
\begin{center}
\begin{tabular}{|c|c|c|c|c|c|c|c|c|c|c|}
\hline
{\bf p1} & {\bf p2} & {\bf p3} & {\bf p4} & {\bf p5} & {\bf p6} & {\bf p7} & {\bf p8} & {\bf p9} & {\bf p10} & {\bf p11} \\
\hline
0 & 0 & 0 & 1 & 1 & 0 & 3 & 0 & 1 & 0 & 1 \\
\hline
\end{tabular} 
\end{center}
\caption{Summary of evaluation of properties for M2MQL.} 
\label{tab:M2MQL}
\end{table}

%------------------------------------------------------------------------- 
\subsection{L5: dmFSQL}

dmFSQL (data mining Fuzzy Structured Query Language)~\cite{20} is an extension of FSQL for data mining. Since FSQL is an extension of the SQL language, then dmFSQL inherits all the properties of SQL. FSQL extends the SQL language with the capabilities to specify flexible queries to address tables that store vague information using fuzzy attributes. dmFSQL introduces a new type of object that does not exist in FSQL. This object holds the initial conditions, intermediate and end results of the DM process. These intermediate results are used to improve the performance of the iterative DM process. 

Advantages of dmFSQL include: (1) similar syntax to SQL; (2) it is understandable; (3) it is possible to express fuzzy conditions in a query; (4) supports iterative DM; and (5) management of any type of data. A drawback of dmFSQL is that it supports only a limited number of DM techniques: clustering and classification. Furthermore, the extensibility of dmFSQL with other DM techniques is considered to be difficult.

\begin{itemize}
\item [] \textbf{Evaluation of Properties}
\item[p1] \emph{Data integration}. There is no evidence in the studied literature that M2MQL supports data integration. 
\item[p2] \emph{Data pre-processing}.	Data pre-processing is unsupported, except for SQL data-set selection function. 
\item[p3] \emph{OLAP}. dmFSQL does not provide constructs for OLAP. 
\item[p4] \emph{Presentation}. Like FSQL, dmFSQL also returns a SQL sentence. They have the same way to present the visualization of results.
\item[p5] \emph{DM methods}. dmFSQL includes two DM methods: clustering and classification. They are often used, but represent only a small part of possible DM methods.
\item[p6] \emph{Implementation}. dmFSQL may be used via the dmFSQL Server that has been programmed mainly in PL/SQL, or a new version of a prototype called DAPHNE~\cite{22} that incorporates the dmFSQL Server.
\item[p7] \emph{Granularity}.	As far as data selection is concerned, because of the properties of SQL, dmFSQL supports operator-level granularity. The DM method in dmFSQL is at task-level.
\item[p8] \emph{Model processing}.	Concerning, model processing, dmFSQL can be reused through the internal invocation: its classification process always needs the prior execution of a clustering process. It does not support model verification, evaluation, validation, or visualization.
\item[p9] \emph{CRISP DM compliance}.	dmFSQL supports partially Data Preparation and Modelling. 
\item[p10] \emph{Background knowledge}.	dmFSQL can not support concept hierarchies. Also there is no functionality for type mapping.
\item[p11] \emph{Extensibility}.	dmFSQL is already a fixed and previously designed extension of FSQL. It is hard to extend. But it is possible, that on the basis of the DM methods in dmFSQL, other DM methods can be described. How to implement this extension is not clear yet.
\end{itemize}

\begin{table}[h]
\begin{center}
\begin{tabular}{|c|c|c|c|c|c|c|c|c|c|c|}
\hline
{\bf p1} & {\bf p2} & {\bf p3} & {\bf p4} & {\bf p5} & {\bf p6} & {\bf p7} & {\bf p8} & {\bf p9} & {\bf p10} & {\bf p11} \\
\hline
0 & 1 & 1 & 1 & 2 & 3 & 2 & 1 & 2 & 0 & 2 \\
\hline
\end{tabular} 
\end{center}
\caption{Summary of evaluation of properties for dmFSQL.} 
\label{tab:dmFSQL}
\end{table}

%------------------------------------------------------------------------- 
\subsection{L6: OLE-DB for DM}

OLE DB is an API comprising SQL capabilities for accessing various data sources. OLE-DB for DM provides the possibility of data mining on OLE-DB providers~\cite{24}. A provider is responsible for the implementation of the OLE-DB interfaces, whereas a consumer accesses data via OLE DB interfaces. Basically, the provider encapsulates the access to data and exposes it to consumers. The interface of OLE-DB for DM that is used for exploring and manipulating the data mining models is similar to the interface that is used for exploring tables. OLE DB for DM represents data mining models as table objects.

\begin{itemize}
\item [] \textbf{Evaluation of Properties}
\item[p1] \emph{Data integration}. As OLE-DB for DM is based on OLE-DB, it provides a rich infrastructure for data integration.
\item[p2] \emph{Data pre-processing}.	Since OLE DB for DM is based on OLE DB, it provides strong capabilities in data selection. It also supports data transformation.
\item[p3] \emph{OLAP}. OLE-DB for DM alone possesses no OLAP capabilities, but can fulfill OLAP requirements via collaboration with OLE-DB for OLAP. OLE-DB for OLAP enables users to perform data analysis via interactive access to various views of the data.
\item[p4] \emph{Presentation}. The DM interface supports model browsing for some models.
\item[p5] \emph{DM methods}. It supports Decision Trees and Clustering.
\item[p6] \emph{Implementation}. Microsoft has implemented an OLE-DB provider for Data Mining based on the OLE-DB for DM Specification. 
\item[p7] \emph{Granularity}. It operates at algorithm level.
\item[p8] \emph{Model processing}. No support for model verification and similar processing tasks; it cannot post-process a model. But, the model can be reused to do prediction.
\item[p9]	\emph{CRISP DM compliance}. OLE-DB for DM (together with OLE-DB) supports Data Understanding, Data Preparation and Modelling. 
\item[p10] \emph{Background knowledge}. There is no support for concept hierarchies. There is no support (also no need) for type mapping in OLE-DB for DM, since OLE-DB performs type mapping before the process enters DM phase.
\item[p11] \emph{Extensibility}. Microsoft data mining provider is open as it is an OLE-DB component. Other DM algorithms can be plugged into the same platform. 
\end{itemize}

\begin{table}[h]
\begin{center}
\begin{tabular}{|c|c|c|c|c|c|c|c|c|c|c|}
\hline
{\bf p1} & {\bf p2} & {\bf p3} & {\bf p4} & {\bf p5} & {\bf p6} & {\bf p7} & {\bf p8} & {\bf p9} & {\bf p10} & {\bf p11} \\
\hline
3 & 4 & 4 & 3 & 2 & 4 & 2 & 2 & 2 & 3 & 3 \\
\hline
\end{tabular} 
\end{center}
\caption{Summary of evaluation of properties for OLE-DB for DM.} 
\label{tab:OLE-DB}
\end{table}

%------------------------------------------------------------------------- 
\subsection{L7: MINE RULE}

The MINE RULE language is an extension of SQL, which enables the extraction of association rules in relational databases~\cite{25}. The user specifies MINE RULE constraints and the format of the association rules that should be extracted. The MINE RULE language may extract the association rules directly from a database. It is also possible to use MINE RULE for rule extraction from data-sets that have previously been prepared for DM. Data selection and integration functionality is provided by the inherent SQL capabilities of MINE RULE. 

Based on MINE RULE, many algorithms and optimization schemas for extraction of association rules have been developed. A design for an association rule-extracting system has also been proposed, although no implementation has been done.

\begin{itemize}
\item [] \textbf{Evaluation of Properties}
\item[p1] \emph{Data integration}. As MINE RULE inherits all SQL capabilities, it provides infrastructure to facilitate data integration inside one database. No effort is made to support data integration between different databases. 
\item[p2] \emph{Data pre-processing}. As MINE RULE inherent all SQL capabilities, it provides the possibility of data selection. However, it supports no normalization, transformation or cleaning functions specific for DM input. 
\item[p3] \emph{OLAP}. There is no interface defined in MINE RULE to support any external OLAP engine. On the other hand, MINE RULE barely supports the features of OLAP without any extension. For example, it does not offer multidimensional views on data; it cannot do statistical aggregation on the data, nor any other complex statistical calculation to support users with multidimensional views on the data. It cannot do time analysis on the data. 
\item[p4] \emph{Presentation}. MINE RULE does not offer the user with any visualization facility. The results from data selection and DM cannot be shown by any means within the context of this language. The results of DM are described as a database table. 
\item[p5] \emph{DM methods}. Association rules are supported.
\item[p6] \emph{Implementation}. No real implementation has been found, although an association rule extraction system has been designed conceptually.
\item[p7] \emph{Granularity}. For dataset selection, MINE RULE reaches operator-level (inherited from SQL). For DM, it behaves at task-level, as it only specifies the task (association rule or cluster or both) but not the algorithm.
\item[p8] \emph{Model processing}. There is no support for model verification, and cannot post-process a model.
\item[p9] \emph{CRISP DM compliance}. MINE RULE supports only Data Preparation and Modelling. 
\item[p10] \emph{Background knowledge}. MINE RULE does not support or recognize concept hierarchies. It offers no means for type mapping in data selection phase or between data selection and DM. According to~\cite{29} MINE RULE provides limited capabilities for background knowledge specification via setting conditions on the values of BODY and HEAD.
\item[p11] \emph{Extensibility}. It is initially designed specifically for the extraction of association rules. But other DM methods can be added by introducing new DM operators. This kind of extension is, however, only possible at a task level.
\end{itemize}

\begin{table}[h]
\begin{center}
\begin{tabular}{|c|c|c|c|c|c|c|c|c|c|c|}
\hline
{\bf p1} & {\bf p2} & {\bf p3} & {\bf p4} & {\bf p5} & {\bf p6} & {\bf p7} & {\bf p8} & {\bf p9} & {\bf p10} & {\bf p11} \\
\hline
1 & 3 & 1 & 1 & 2 & 2 & 3 & 0 & 2 & 0 & 2 \\
\hline
\end{tabular} 
\end{center}
\caption{Summary of evaluation of properties for MINE RULE.} 
\label{tab:MINE}
\end{table}

%------------------------------------------------------------------------- 
\subsection{L8: Oracle Data Mining}

Oracle Data Mining (ODM)~\cite{31} offers a data mining development and deployment platform inside Oracle databases, along with a GUI named Oracle Data Miner. The user uses the GUI to navigate through the process of creating, testing, and applying models – approximating the CRISP-DM methodology. ODM provides a PL/SQL package to create, destroy, describe, apply, test, export and import models. According to METAspectrum's evaluation report on data mining tools, ODM is in the leader segment together with SAS and SPSS. Some of the interesting features of ODM are:

\begin{itemize}
\item Automatic data preparation for each model.
\item It supports the Java standard for data mining (JSR-73). 
\item Bioinformatics Sequence Search and Alignment (BLAST) available in the Oracle Database via table functions.
\end{itemize}

\begin{itemize}
\item [] \textbf{Evaluation of Properties}
\item[p1] \emph{Data integration}. Data integration is supported via Oracle Gateway technology. 
\item[p2] \emph{Data pre-processing}. ODM provides utilities for data preparation steps prior to model building such as outlier treatment, discretization, and normalization. 
\item[p3] \emph{OLAP}. Support for OLAP is available via another option for Oracle Enterprise Edition, namely Oracle OLAP DML.
\item[p4] \emph{Presentation}. Presentation is supported via ODM and MS Excel Add-In.
\item[p5] \emph{DM methods}. Since Oracle version 11g, ODM supports data transformation and model analysis, classification, regression, clustering, association rules, feature extraction as well as text and spatial mining.
\item[p6] \emph{Implementation}. The platform has been fully implemented in Oracle RDBMS.
\item[p7] \emph{Granularity}. ODM provides coarse grained granularity on SQL function call level.
\item[p8] \emph{Model processing}. Model processing is supported.
\item[p9] \emph{CRISP DM compliance}. The following phases are supported via the platform inside the Oracle RDBMS and ODM: data understanding, data preparation, modelling, evaluation and deployment.
\item[p10] \emph{Background knowledge}. We found no clear statement in the literature that background knowledge is supported. But, considering the provided support for other properties, we may infer that background knowledge should be supported; other properties could not be supported to this extent without the background knowledge.
\item[p11] \emph{Extensibility}. The data mining package can be extended via additional methods.
\end{itemize}

\begin{table}[h]
\begin{center}
\begin{tabular}{|c|c|c|c|c|c|c|c|c|c|c|}
\hline
{\bf p1} & {\bf p2} & {\bf p3} & {\bf p4} & {\bf p5} & {\bf p6} & {\bf p7} & {\bf p8} & {\bf p9} & {\bf p10} & {\bf p11} \\
\hline
4 & 3 & 4 & 4 & 4 & 5 & 1 & 4 & 3 & 3 & 3 \\
\hline
\end{tabular} 
\end{center}
\caption{Summary of evaluation of properties for Oracle Data Mining.} 
\label{tab:Oracle}
\end{table}

%------------------------------------------------------------------------- 
\section{Evaluation Summary}
\label{sec:summary}

\begin{figure*}
	\begin{center}
		\includegraphics[scale=0.325]{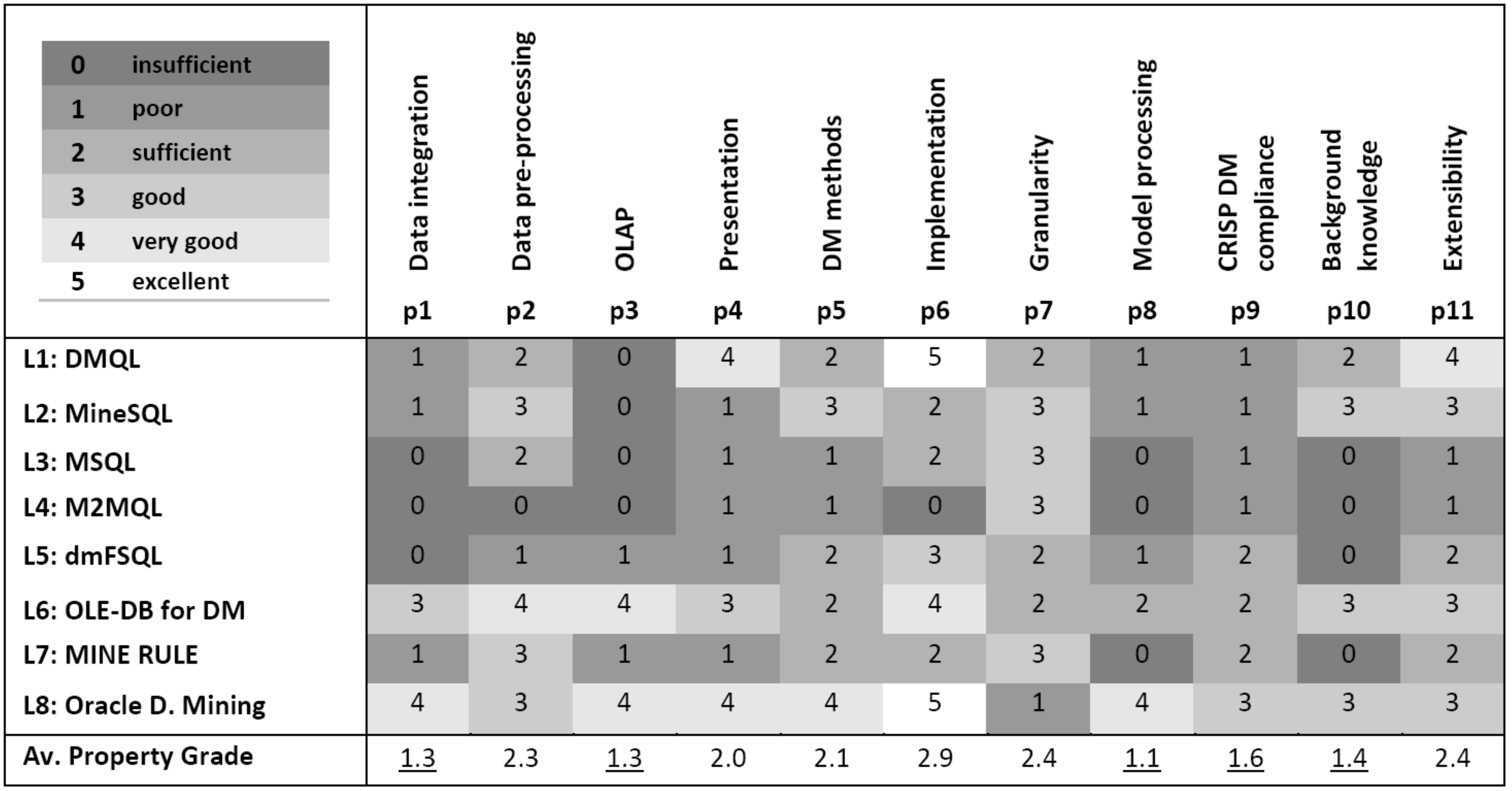}
	\end{center}
\caption{Evaluation summary. Most of the surveyed languages provide less than sufficient support for data integration, OLAP, model processing, CRISP-DM compliance, and background knowledge.} 
\label{fig:evaluation_summary}
\end{figure*}

In Figure~\ref{fig:evaluation_summary}, dark areas represent lack of the adequate support, whereas bright areas indicate excellent support of evaluated languages for properties. 

Properties where the average grade is less than 2.0 (that is less than sufficient) are: \emph{data integration}, \emph{OLAP}, \emph{model processing}, \emph{CRISP-DM compliance}, and \emph{background knowledge}. For these properties most of the surveyed languages provide less than \emph{sufficient} support. The average grade for a property $p_{i}$ is calculated as follows: 

\begin{center}
$\frac{\sum_{j=1}^{8} G(p_{i},L_{j}) }{8}$
\end{center}

Table~\ref{tab:ranking} shows the ranking of the surveyed languages based on average grades. The average grade for a language $L_{j}$ is calculated using  this formula: 

\begin{center}
$\frac{\sum_{i=1}^{11} G(p_{i},L_{j}) }{11}$
\end{center}

Oracle Data Mining received the highest average grade (that is 3.5), whereas M2MQL the lowest one (that is 0.6).

\begin{table}[h]
\begin{center}
\begin{tabular}{|l|l|c|}
\hline
{\bf Rank} & {\bf Language} & {\bf Average Grade} \\ 
\hline
1 & L8: Oracle DM & 3.5 \\ 
2 & L6: OLE-DB for DM & 2.9 \\ 
3 & L1: DMQL & 2.2 \\ 
4 & L2: MinieSQL & 1.9 \\ 
5 & L7 MINE RULE & 1.5 \\ 
6 & L5: dmFSQL & 1.4 \\ 
7 & L3: MSQL & 1.0 \\ 
8 & L4: M2MQL & 0.6 \\
\hline
\end{tabular} 
\end{center}
\caption{Language ranking based on average grades.} 
\label{tab:ranking}
\end{table}

%------------------------------------------------------------------------- 
\section{Conclusions and Future Work}
\label{sec:conclusions}

We have presented a survey of the state-of-the-art of DMI-related languages. For each surveyed language we provided an overview and stated the interesting features and drawbacks that we have observed. 

For the purpose of systematic language evaluation, we have developed a list of properties (or criteria) and a scoring system. For each property we have provided a single paragraph description on how well the language under this study supports the property. We used our scoring system to quantify our opinion. We have observed that most of the surveyed languages provide less than sufficient support for data integration, OLAP, model processing, CRISP-DM compliance, and background knowledge.

In future, based on our observations made during this survey, we plan to propose a model and the corresponding language that will advance the state-of-the-art in DMI languages.

%------------------------------------------------------------------------- 
\section*{Acknowledgment}

The research leading to these results has received funding from the European Union Seventh Framework Programme (FP7/2007-2013) under grant agreement 215024 (ADMIRE Project, www.admire-project.eu).

%------------------------------------------------------------------------- 

%------------------------------------------------------------------------- 
\end{document}